\documentclass[manuscript]{aastex}
\begin{document}
\title{High resolution imaging of the ATLBS regions: the radio source counts}
\author{K. Thorat\altaffilmark{1}}
\affil{Raman Research Institute, C. V. Raman Avenue, Sadashivanagar, Bangalore 560080, India}
\email{kshitij@rri.res.in}
\altaffiltext{1}{Joint Astronomy Programme, Indian Institute of Science, Bangalore 560012, India}
\author{R. Subrahmanyan}
\affil{Raman Research Institute, C. V. Raman Avenue, Sadashivanagar, Bangalore 560080, India}
\author{L. Saripalli}
\affil{Raman Research Institute, C. V. Raman Avenue, Sadashivanagar, Bangalore 560080, India}
\author{R. D. Ekers}
\affil{CSIRO Astronomy and Space Sciences, Epping, NSW 2121, Australia}

\begin{abstract}

The Australia Telescope Low-brightness survey (ATLBS; \cite{SESS10})
regions have been mosaic imaged at a radio
frequency of 1.4~GHz with $6\arcsec$ angular resolution and
72~$\mu$Jy~beam$^{-1}$ rms noise.  The images (centered at 
RA: $00^{h}\ 35^{m}\ 00^{s}$, DEC: $-67^{\circ}\
00^{'}\ 00^{''}$ and RA: $00^{h}\ 59^{m}\ 17^{s}$, DEC: $-67^{\circ}\
00^{'}\ 00^{''}$ (J2000 epoch)) cover 8.42 square degrees
sky area and have no artifacts or imaging errors above the image thermal
noise. Multi-resolution radio and
optical r-band images (made using the 4-m CTIO Blanco telescope) were used
to recognize multi-component sources and prepare a source list; the
detection threshold was 0.38~mJy in a low resolution radio image made with beam FWHM of
$50\arcsec$. Radio source counts in the flux density range 0.4-8.7~mJy are estimated, with
corrections applied for noise bias, effective area correction and
resolution bias. The resolution bias is mitigated 
using low resolution radio images, while
effects of source confusion are removed by using high resolution images
for identifying blended sources. Below 1~mJy the ATLBS counts are systematically lower than the previous estimates. Showing no evidence for an upturn down to 0.4 mJy, they do not require any changes in the radio source population down to the limit of the survey. The work suggests that  
automated image analysis for counts may be dependent on the ability of the 
imaging to reproduce connecting emission with low surface brightness and on the 
ability of the algorithm to recognize sources, which may require that source 
finding algorithms effectively work with multi-resolution and multi-wavelength data. The work underscores the importance of using source lists---as opposed to component lists---and correcting for
the noise bias in order to precisely estimate counts close to the image
noise and determine the upturn at sub-mJy flux density.

\end{abstract}
 

\section{Introduction}
\label{section_introduction}
Historically, radio source counts has been a key observational probe of cosmology, more specifically of the geometry of the universe. In a Euclidean universe the volume scales with distance as $V \propto r^{3}$ whereas the flux density scales as $S \propto r^{-2}$, which means that the integral number of sources (for a population of non-evolving sources with constant comoving number density) above any flux density scales as $n \propto S^{-3/2}$. Departures from this expectation was key evidence for  non-Euclidean geometry.  More recently, the geometry of the cosmos has been established with precision, and source counts represent a measure of cosmological evolution in radio source populations. The behavior of counts at sub-mJy flux density, the nature and evolution of these sources and the question of whether they constitute a new population are unclear.\newline

At flux densities of $\leq 1.0$~mJy a `flattening' of normalized differential source counts has been widely reported in literature (Windhorst et al. 1985 (WMO85 henceforth), \cite{Hop03}, \cite{HJNP05} and references therein). The rms noise level in the latter studies are: $45\ \mu$Jy for WMO85, varying between $12$ to about $100\ \mu$Jy for Hopkins et al. (2003) (see Fig.9 from Hopkins et al. (2003)) and approximately $10\ \mu$Jy for Huynh et al. (2005). The flattenning is observed as an apparent change of slope from $\sim0.7$ at $5.0 - 100.0$ mJy to about $~ 0.4$ in the $0.25- 5.0$ mJy range. WMO85 examined the optical identifications of faint radio sources and found that the sub-mJy radio source population is dominated by blue spiral galaxies.  Later studies, however, have arrived at discordant results and identify the sub-mJy sources with different populations: starburst galaxies (\cite{Con89}, \cite{Ben93}, \cite{HJNP05}), early type galaxies \citep{GMZ99}, low (radio) luminosity active galactic nuclei (AGNs) \citep{HJNF08} or a mixture of these. Since spectroscopically complete samples of sub-mJy sources are not available, the exact nature of the population observed as  sub-mJy radio sources remains uncertain; however, it is widely agreed that flattening below $1.0$ mJy requires an evolving population that is different from those that dominate counts at higher flux densities.

It may be noted here that the literature is not consistent in observing a flattening in counts at sub-mJy flux density; for example, the counts in \citet{Pra2001} and \citet{SESS10} are consistent with a continuation in the slope of the differential counts below mJy flux density.  Potential causes for the discrepancy are that deep radio surveys often suffer from inadequate sky coverage that is
necessary to average over clustering, wide-field surveys often do not go sufficiently deep and so biases arising from the
proximity of the detection threshold to the image noise may distort measured counts.  Additional uncertainly arises from lack of
understanding of the radio structures in sub-mJy radio sources, which may lead to biases related to the ability of the survey to
catalog sources from detections of source components.\newline

The ATLBS is a moderately wide-field radio survey, covering $8.42$ square degrees in the southern sky at $1.4$ GHz. The survey has been carried out using the Australia Telescope Compact Array (ATCA), which is a Fourier imaging interferometer array. The radio observations for the low-brightness survey were designed for complete u-v coverage up to $750$~m (for details of these observations, see \cite{SESS10}).  The work presented herein is based on the combination of the initial low resolution survey and  more recent radio observations with the ATCA in multiple expanded array configurations that extended the u-v coverage to $6$~km, giving a synthesized beam with FWHM of $6\arcsec$ and images with rms noise $72\ \mu$Jy~beam$^{-1}$. The high resolution observations are critical in better estimation of the source structures and for removing the effects of blending(source confusion) in the low resolution survey. See \cite{SST11} for details on a sample of extended sources and classification of sources from the ATLBS survey, which has been carried out using both the high resolution and low resolution images.

The improved imaging of the ATLBS survey regions has been used to revisit the 1.4~GHz source counts: the survey has sufficient sensitivity to probe the sub-mJy regime, and the relatively large sky coverage avoids clustering related uncertainties.  A specific improvement in this work is the care taken to identify sources with low surface brightness by making use of low resolution images, and using multiple indicators to identify components of sources. The low resolution images were used to make initial identifications so that the sources which might be resolved into multiple components at higher resolution are identified correctly. The blending issues inherent in using low resolution images have been avoided using higher resolution images to identify blends. In addition, the use of low resolution images (beam FWHM = $50\arcsec$) almost completely removes effects of resolution bias (for a detailed discussion, see Section.~\ref{section_resolution_bias}). These strategies, together with use of optical images\footnotemark  to locate candidate galaxy hosts and a careful visual examination of resolved and complex sources
instead of automated classification ensures that the ATLBS catalog is a `source catalog' as opposed to a `component catalog'. The distinction between `sources' (which are single sources) as opposed to components (which may be parts of a single source or unrelated sources which are close to each other due to projection effects) is crucial in estimating the true source counts. \footnotetext[1]{The optical images used to identify the candidate hosts have been made using the CTIO 4-meter Blanco telescope, in SDSS ${\rm r}^{\prime}$ band. The optical images cover most of the regions A and B. A detailed description of the optical imaging will be presented in \cite{TSS12}.}\newline

The organization of the paper is as follows. In the next section, we give details of the new high resolution radio observations of the ATLBS survey regions. In Section (\ref{section_source_detection}) we give details of the procedure adopted for source detection and the estimation of source flux densities. In Section (\ref{section_source_counts}) we present the source counts along with the corrections necessary to derive the true counts from the observations. Next, in Section (\ref{section_comparison}), we compare the ATLBS source counts with previous work. Finally, in section (\ref{section_conclusions}), the conclusions are presented.

\section{Observations}
\label{section_observations}
The radio observations which form the ATLBS survey were made with different array configurations of the ATCA. The observations were made in the $20$~cm band, with a center frequency of $1388$~MHz. These observations recorded visibilities in full polarization mode. The observations were made in two sub-bands with center frequencies of $1344$ MHz and $1432$ MHz, with bandwidths of $128$~MHz. Each band was covered by $16$ independent frequency channels of which multi-channel continuum visibilities in the 13 central channels were used. \newline

The ATLBS survey covers two regions in the southern sky. These regions are designated as `A' and `B' having their centers at RA: $00^{h}\ 35^{m}\ 00^{s}$, DEC: $-67^{\circ}\ 00^{'}\ 00^{''}$ and RA: $00^{h}\ 59^{m}\ 17^{s}$, DEC: $-67^{\circ}\ 00^{'}\ 00^{''}$ (J2000 epoch) respectively. Together they cover an area of $8.42$ square degrees. The two regions were specifically selected to be devoid of strong radio sources, low 
Galactic foreground emission, and at an optimum latitude that allowed for good visibility coverage for radio observations with the ATCA (see \cite{SESS10} for details). Each of the regions is covered in the interferometer observations as a mosaic of $19$ pointings.\newline

Earlier observations were made with the array configurations $750$A, $750$B, $750$C and $750$D. These configurations provided complete u-v coverage up to $750$ meters, and low-resolution images made with these data were presented in \citet{SESS10}. To obtain good u-v coverage up to $6$ km antenna separations, new observations were made with the array configurations $6$A, $6$B, $6$C, and $6$D. The plan of the observations was as follows. The two regions A and B were observed for $12$ hours with the array configured in each of the four layouts. The total observing time was $96$ hours and this was spread over the mosaic pointings. Each pointing was observed for $20$ seconds before switching to the next pointing and the pointings were thus cycled over every $19 \times 20$ sec.\newline

The reduction and imaging of the data was done with the radio interferometer data reduction software MIRIAD \citep{STW95}.
The data was calibrated for amplitude and phase using the calibrator PKS B2353-686. The absolute flux scale was set using the calibrator PKS B1934-638. The visibility data was also examined for radio frequency interference and outliers and other obviously corrupted data were rejected.  First, images were produced with the $6$ km configuration visibilities.  Deconvolution used the Clark algorithm \citep{clark}, which is faster for large images. After a phase-only self-calibration iteration, the clean components of sources exclusively within the primary beam main lobe were selected by masking the regions outside the primary beam main lobe in the images.  Visibilities corresponding to these were subtracted from the data and the residual visibilities were imaged to model the sources outside the primary beam main lobe.   The model was allowed to include both positive and negative intensity components since the source structures in these regions were unreal because of the significant azimuthal structure in the telescope primary beam sidelobes, which modulated the visibilities of sources outside the primary beam as the alt-azimuth antennas tracked the pointing centers over the observing session. Visibility domain subtraction of this model provided a dataset containing essentially only  sources inside the primary beam main lobe.  Next, the visibilities from the array configurations of 750~m length were reduced in a similar way to get visibilities referring only to the sources within the primary beam main lobe.  Following these steps, the visibilities were all concatenated for joint imaging, deconvolution and iterative self-calibration.

\begin{figure}
\includegraphics[width = 3 in, angle = 270]{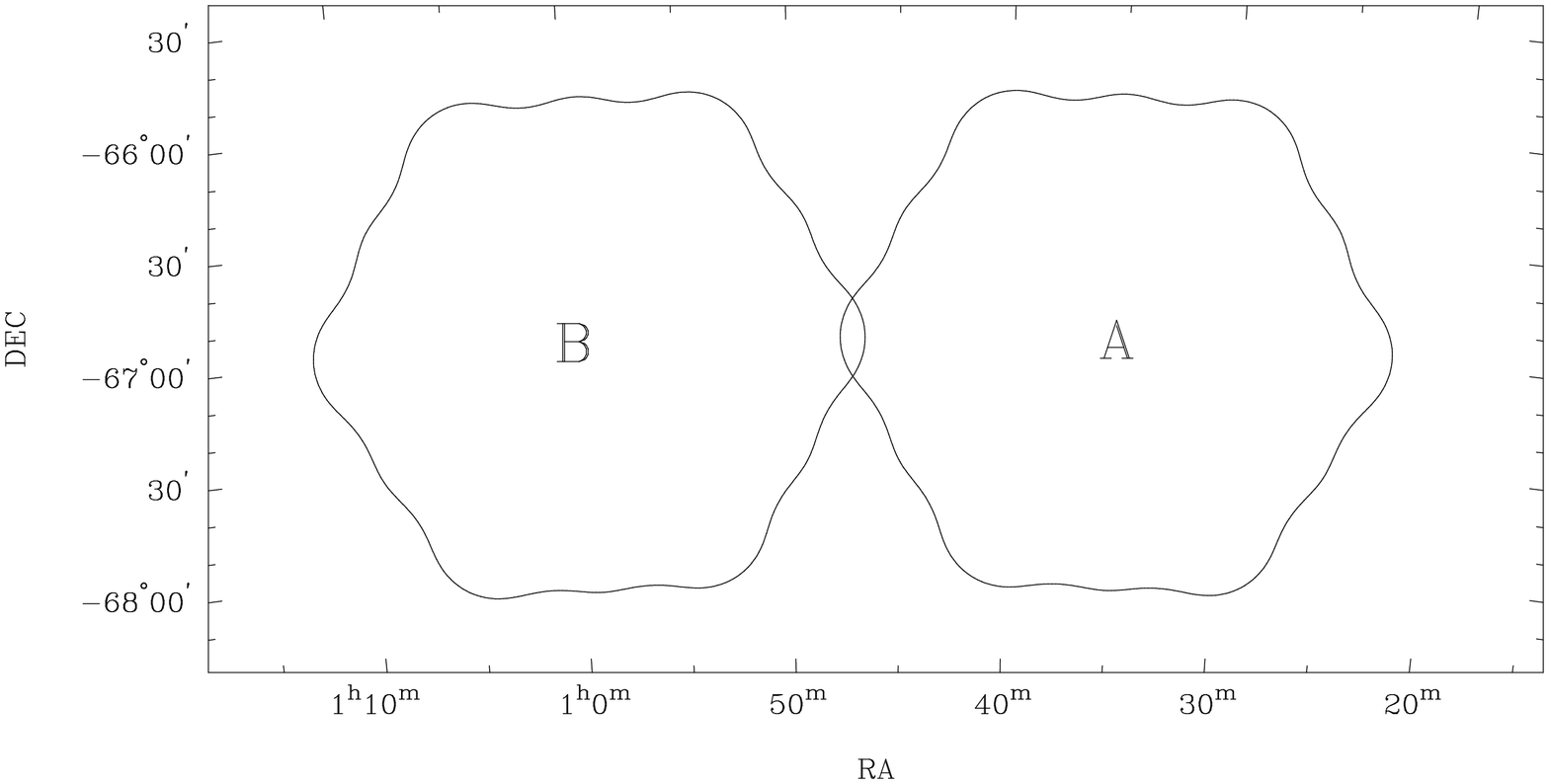}
\caption{The boundaries of regions A and B are shown. }
\label{fieldab}
\end{figure}

\begin{figure}
\includegraphics[width = 5 in,angle = 270]{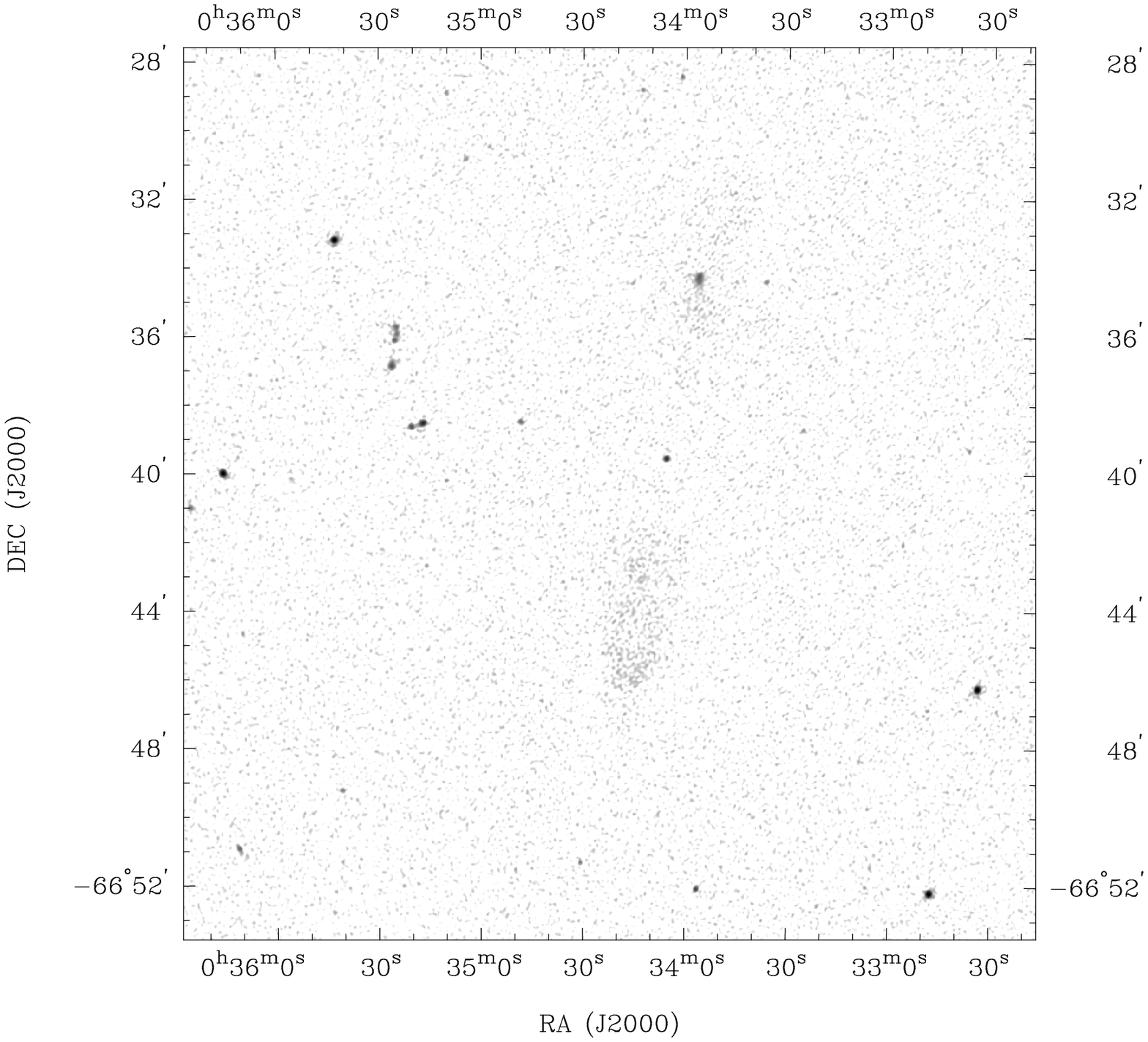}
\caption{A representative region from Field A shown in greyscale. The image has beam FWHM $6\arcsec$ and rms noise of $72\ \mu$Jy~beam$^{-1}$. }
\label{fielda}
\end{figure}

\begin{figure}
\includegraphics[width = 5 in, angle = 270]{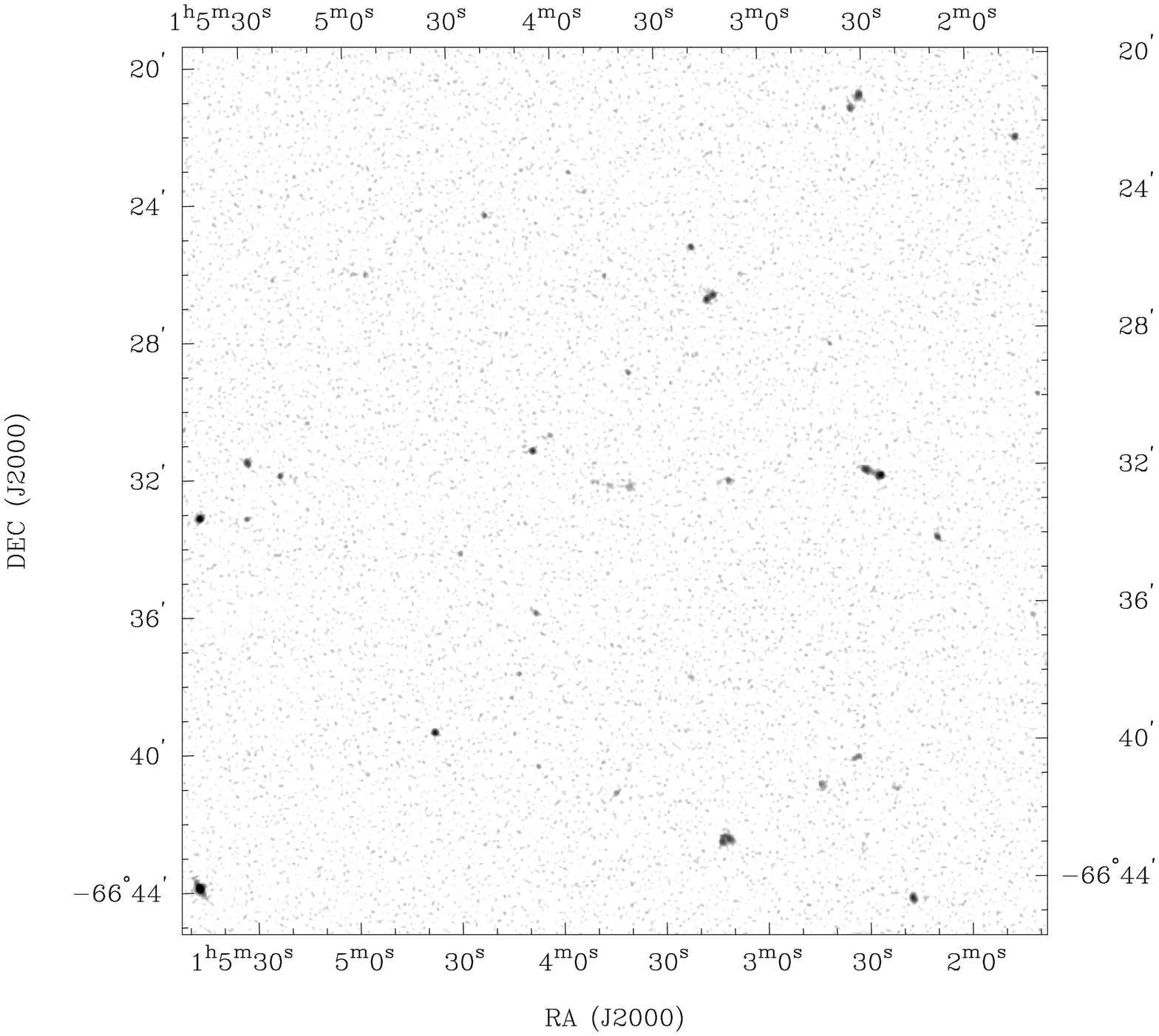}
\caption{A representative region from Field B shown in greyscale. The image has beam FWHM $6\arcsec$ and rms noise of $72\ \mu$Jy~beam$^{-1}$. }
\label{fieldb}
\end{figure}

\begin{figure}
\includegraphics[width = 6 in]{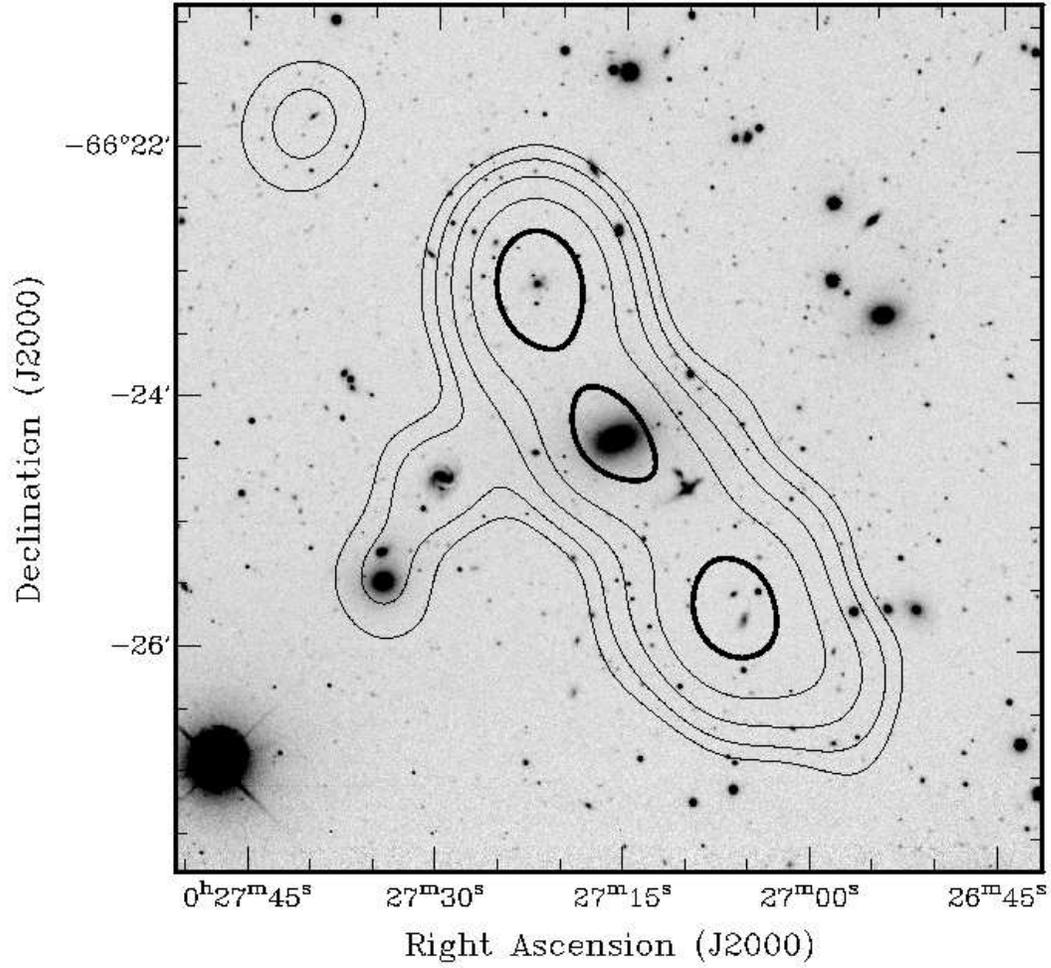}
\caption{A sky region from low resolution image. The countours are selected to increase in factors of 2, starting from $0.38$ mJy. The greyscale depicts the optical ${\rm r}^{\prime}$ band image derived from our observations using the MOSAICII imager on the CTIO 4-meter telescope \citep{TSS12}. The image reaches a
depth of ${\rm r}^{\prime}$= 22.5. }
\label{lowres_source}
\end{figure}

\begin{figure}
\includegraphics[width = 6 in]{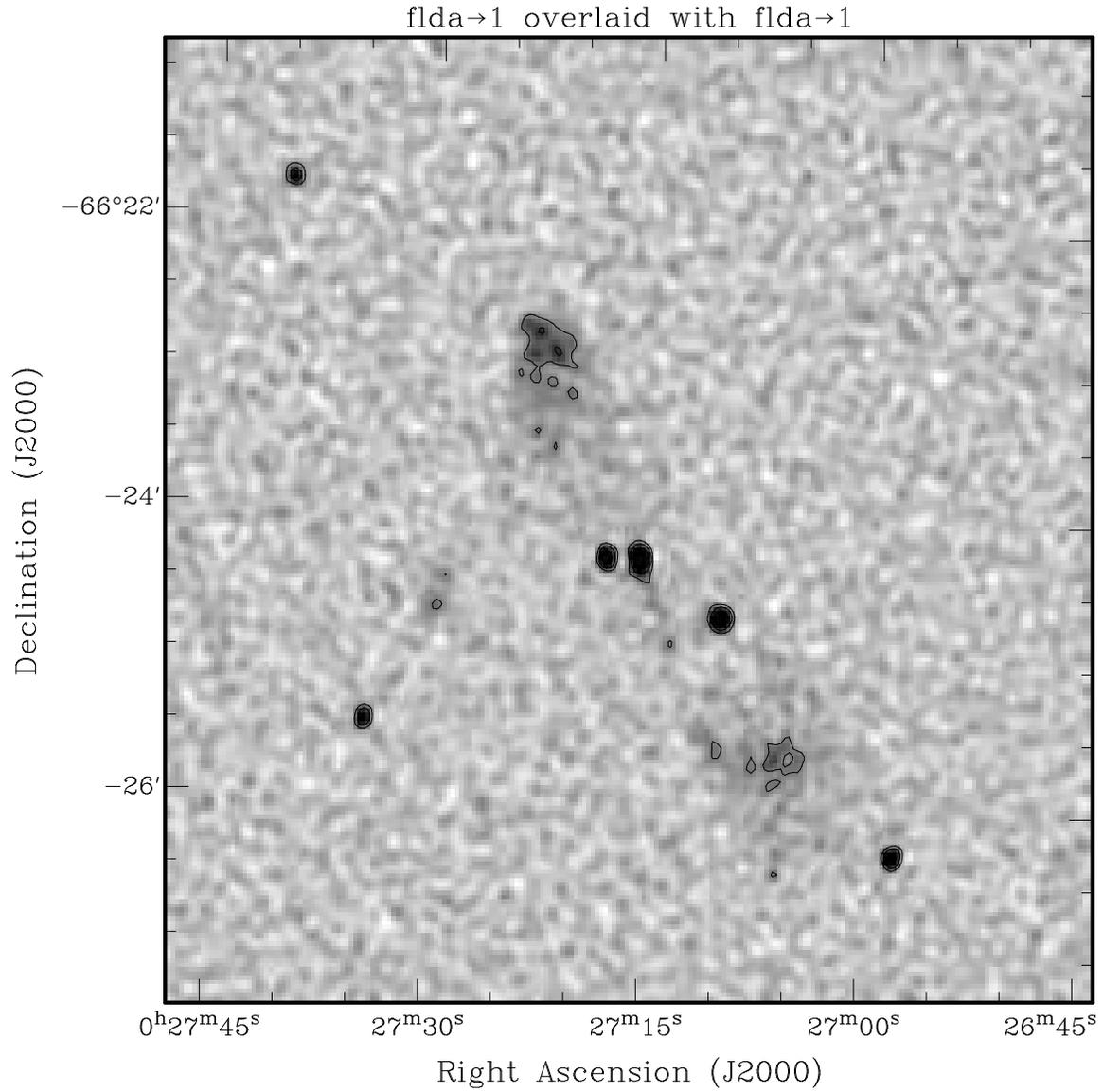}
\caption{The same region as in Fig.~\ref{lowres_source}, shown in high resolution. Note that what may be perceived as a single  source at low resolution is a composite of multiple sources. For a discussion of blending-confusion issues see \ref{section_blending_confusion}. The octave countours start from $0.28$ mJy.}
\label{highres_source}
\end{figure}

The imaging was done with large sidelobe suppression area to obtain a good beam.  Initial self calibration iterations were of phase-only type. The multi-frequency deconvolution described in \cite{sault} was used to get the clean components from self-calibrated data.  Because the u,v-coverage was not complete in the 750-6000~m range, deconvolution was assisted by masking areas of the image where no significant intensities were present. This was done by first smoothing the image to detect most of the diffuse emission and in the smoothed image regions below $4.5\ \sigma$ were masked. This process was repeated, iteratively, along with phase and amplitude self calibration. The solution intervals for the successive iterations varied from $15$ to $5$ minutes.  In a final step, the images of individual pointings were regridded and stitched together to obtain the combined mosaic image of region A and B separately. The sky area covered by the  by these regions is depicted in Fig.~\ref{fieldab}.  Sample sub-images of these high resolution images of these regions are presented in Fig.~\ref{fielda} and Fig.~\ref{fieldb}. To better appreciate the difference between the high and low resolution images, we have presented a sky region in low and high resolutions in Fig.~\ref{lowres_source} and Fig.~\ref{highres_source}. The linear combination of pointings was done with a weighting that maintained rms noise in the images nearly uniform; this value is $72\ \mu$Jy~beam$^{-1}$.  Over the 8.42 square degrees sky area where the gain in the mosaic images exceeds 0.5 (and hence the rms noise in primary beam corrected images does not exceed the above value by more than a factor of two) there are no image errors apparent above the thermal image noise.  This exceptional quality in the wide-field images makes it a useful database for automated source finding and classifying algorithms as well as reliable studies of radio source properties.
In addition, the range of source morphologies makes possible the validation and developement of automated source finding algorithms, which may become crucial in the analysis of data from large surveys such as LOFAR and SKA.

\section{On the identification of sources in the ATLBS survey}
\label{section_source_detection}

We have jointly used the low and high resolution ATLBS images in the strategy adopted for source detection and classification as well as for estimations of flux density. The former, with $50\arcsec$ FWHM beam, was made using the 750-m arrays and the latter, with $6\arcsec$ FWHM beam, was made using all of the visibilities up to 6~km baselines. The low resolution images are used to identify sky regions in which to search for source components using the high resolution images.   The low resolution images have relatively lower chance of missing a source due to its being resolved: known in the literature as `resolution bias', resolved sources with same integrated flux density would have relatively lower peak flux density and hence may not have image pixels above the detection threshold; this would result in high resolution images missing such sources that may be detected in lower resolution images.  The high resolution images have been used to detect source components, identify blends and classify the sources. Various aspects that potentially introduce bias and errors in source detections such as the thermal noise in the images and the effects of blending-confusion have been examined, as described below.

\subsection{Thermal Noise Considerations}

As mentioned above, we identify regions in the low resolution image which contain sources. The size of these regions, or `footprints' are of importance in determining the threshold for source detection, along with considerations related to the image thermal noise (where `thermal' noise is the composite of the receiver noise, the sky as well as ground noise) and source blending-confusion effects in the high resolution image. We describe here the procedure to determine the detection thresholds separately for source identification in low resolution images and for component identification in high resolution images.\newline

A source detection threshold can be derived for an image if the number of spurious sources expected above any threshold may be estimated. The latter can be calculated easily if the noise follows a Gaussian distribution with a known variance. But for an image convolved with a point spread function, this would only give the total area in the image above the threshold. The actual number of spurious `sources' or `islands'(connected pixels) above the threshold would depend on the shape and size of the point spread function. Above the threshold only the summits of the spurious `sources' might be observed, with a smaller area than that corresponding to the FWHM of the main lobe of the point spread function. Therefore, the detection threshold needs to be determined taking this into account.  \newline

\subsubsection{Detection Threshold in the Low Resolution Image}
For the low resolution image, we employ an empirical approach towards threshold determination. Since the initial noise distribution before being convolved with the point spread function is expected to be Gaussian (with a zero mean), we expect the number of positive noise peaks above a flux density value (say $S_{cutoff}$) to be the same as the number of negative noise peaks below $-S_{cutoff}$. Since the positive noise peaks cannot be distinguished from real sources, we use the number of negative peaks to estimate the detection threshold. The number of negative peaks is affected by the area which is covered by `sources' at every detection threshold. This is because a negative peak occurring at the location of a source will not be detected as a negative peak in the counting. We correct for this by the ratio of total area covered by the sources above any threshold and the total image area. For example, we find $68$ noise peaks below $-0.38$~mJy in the survey region and, after correction for the above, we estimate that there are $71$ spurious noise peaks below this threshold in the entire survey region including those areas covered by genuine sources. We expect very similar numbers of positive noise peaks above $+0.38$~mJy in the survey region. If we assume Gaussian statistics  we expect only about 2 spurious sources in the entire survey region above a $4 \sigma$ threshold; the significantly larger number of spurious sources suggests that the image noise is non-Gaussian at this level. Since we find $1244$ sources above $0.38$~mJy in the ATLBS survey region, we conclude that the fraction of spurious sources at a threshold of $0.38$ mJy is $5.7\%$. The percentage of spurious sources changes with the detection threshold. The percentage of spurious sources goes from $41\%$ at $0.29$~mJy to merely $0.3\%$ at a threshold of $0.48$~mJy. However, the number of sources which will not be detected will also rise with the threshold. We therefore choose $0.38$ mJy as our threshold for the low resolution image (a $4 \sigma$ threshold), with an expected $5.7\%$ of false detections. Therefore the reliability of the catalog is $94.3 \%$ at the detection threshold. The number of negative peaks might be affected by the fact that negative peaks have not been deconvolved, and they would retain the dirty beam pattern. This may give rise to negative peaks arising from the sidelobes of the beam. However, since the sidelobe level is below $10\%$, the contribution from the sidelobes to negative peaks is not expected to affect the above analysis. It may be noted here that the above discussion does not consider the sources which have true flux density below the detection threshold but are detected above the threshold due to noise peaks biasing their intensity above the threshold.  We have carried out simulations of noise bias which estimates this effect and we correct the source counts for this effect; these corrections are discussed below in section (\ref{section_source_counts}).  \newline   

\subsubsection{Detection Threshold in the High Resolution Image}
For the high resolution image, we follow a different strategy for determining the thermal noise limited detection threshold. Because we only search for the high resolution peaks in the footprints of the sources detected in the low resolution images, the threshold for the detection of the high resolution peaks is determined using the footprint area in which the source components might be located and not the complete area of the image. In this way, we may make use of a unique detection threshold for each source component which would depend on the footprint area in which the source component is located. We aim to keep the number of spurious components detected in high resolution images to be one per $10$ footprints. Since in most cases one footprint corresponds to one source detected in the high resolution image, the above criterion corresponds to a false detection rate of approximately $10\%$. We empirically determine the mean number of pixels per negative peak by measuring the number of negative peaks and pixels at a given flux density cutoff. We find  that the number of pixels per peak is $1.37$ for specifically the synthesized beam and image sampling in the high resolution ATLBS image. Assuming that the number of spurious positive pixels is similar to that of negative pixels, the number of spurious positive peaks can be estimated. For each footprint (in which we detect source component peaks in the high resolution image) we choose the flux density threshold so as to keep the probability of including noise peaks as spurious source components below $10\%$. If we choose a low threshold for identifying peaks in the high resolution image as source components, then a large number of spurious components would be considered as source components and, therefore, a large number of sources would be misclassified as multicomponent and hence complex. If we choose a high threshold, then genuine source components might be missed and sources would be misclassified as extended and resolved in the high resolution images or misclassified as unresolved when they are actually complex. However the latter misclassification is avoided by recognizing that the integrated flux density in the high resolution image falls short of that in the low resolution image and by classifying such sources as complex in these cases. Misclassification of unresolved sources would happen predominantly in cases where the peak flux density is close to the rms image thermal noise. For a flux density threshold of $S_{th}$, if the probability of occurrence of spurious pixels above the threshold is $P(S_{th})$ (the probability is determined by integrating the assumed Gaussian distribution for the pixels), then the number of spurious source components in the footprint area is $A_{f}\ P(S_{th})/1.37$, where $A_{f}$ is the number of pixels in the footprint. Since we wish to keep the number of spurious source components in each footprint less than $0.1$, the flux density threshold is given by $P(S_{th})< 1.37 \times 0.1/(A_{f})$. We have used this recipe to compute the flux threshold uniquely for each footprint. As we do not directly estimate the number of spurious sources from the number of negative peaks in the image, the effect due to the sidelobes of the dirty beam, as descibed above, does not influence the detection threshold directly.\newline

The thermal noise limits discussed above along with considerations related to source blending-confusion, which are discussed next, lead to the adopted detection threshold for the ATLBS survey and hence determine the error in source counts and confidence in classification.

\subsection{Source Blending-Confusion}
\label{section_blending_confusion}

Source blending-confusion (also called source confusion in literature) is because any telescope of finite size has only a finite angular resolution. Since the synthesized beam has a finite size, this makes the detection of two distinct sources that are located close to each other difficult because the sources may appear blended with each other in the image. Blending-confusion limits the flux density limit to which sources may be reliably identified distinctly in a survey, this is a limitation apart from that arising due to the image thermal noise. See \citet{Jaun68} and \citet{MC85} for some of the earliest descriptions of this effect. In this section we describe the effects of the source blending-confusion on source detection in the ATLBS survey. Since thermal noise considerations suggest a source detection threshold of $0.38$ mJy for the low resolution image, we examine the source blending-confusion for this threshold to confirm that the latter are sub-dominant to the limitation arising from thermal noise. \newline

We estimate the effects of source blending-confusion for the ATLBS high resolution images (beam FWHM = $6\arcsec$) using simulations, which we describe next. We assume that the flux density distribution of sources follow the source counts derived from the PDS survey (Hopkins et al., 2003). The simulation Poisson distributed sources over a sky area corresponding to the total area imaged in the ATLBS survey ($8.42$  square degrees). Blending-confusion causes sources to be shifted from the bin corresponding to their true flux density to a bin with higher flux density. The source counts in any given flux density bin change due to the migration of sources in and out of the bin: into the bin from lower flux density bins and out of the bin to  higher flux density bins. The flux density range for the simulation was chosen to accurately model the effects of blending- confusion for our survey. The simulations allow for multiple blendings with other sources. The lower limit of the range was chosen so that the effects of the sources changing bins due to blending-confusion are reliably modeled for the bins above $0.38$ mJy, which is our detection threshold flux density. The upper limit of the range is chosen so that the brightest sources in the survey are accounted for (in the simulation, we expect to detect only a single source in the highest flux density bin).  Additionally, the shapes of the sources in the synthesized images (which have been restored following deconvolution) were set to be Gaussian with a FWHM of $6\arcsec$. The `radius' of each source was taken to be the distance from the source peak at which the intensity drops to half the threshold flux density. Sources were counted as blends if the distance between their peaks was smaller than the sum of the radii of the two sources, since in this situation the two sources would appear to be connected by emission above the threshold flux density in the image. Those sources that have peak flux density less than half the threshold flux density would appear `blended' if they are located within the radius of a source that has peak exceeding half the threshold flux density, {\it i.e.}, if the radius of the stronger source is smaller than the distance between the peaks. An arbitrarily small radius $\ll 6 \arcsec$ is assigned to such sources, which allows for the possibility that two such sources blend together and then blend with other sources.  \newline

The simulations were made with octave bin ranges from 0.1~mJy upwards, and showed that the recovered flux density distribution  was not significantly altered. Most blending-confusion of sources occurs at the lowest flux density bins, and the source counts remain unchanged for higher flux density bins. The number of sources migrating into a bin from lower flux density bins and the number of sources migrating out of the bin due to blending-confusion were also estimated. In the bin $0.4 - 0.8$ mJy the source distribution was unchanged within errors.  For these and higher flux density bins, the number of sources migrating to and out of the bin matches. For sources in bins above a flux density of $3.2$ mJy, the blending-confusion effects were found to be negligible, possibly due to the relatively sparse distribution of sources above this flux density.  For source counts derived herein from the ATLBS, based on a detection threshold of $0.38$ mJy, the correction factors required to account for source blending-confusion is less than $0.1\%$. \newline

We infer from the above analysis that the threshold for source detection in the high resolution images might be placed at or even below $0.38$ mJy, insofar as blending-confusion is concerned, while keeping the errors introduced in the source counts due to blending-confusion below $1\%$. Therefore, we may choose the detection threshold for components in the high resolution image (within the footprint area as outlined in the previous section) based solely on thermal noise effects in the high resolution image. We note here that it is possible that the low resolution image may have blends, which will be identified as separate sources using the high resolution image.

\subsection{Bandwidth Smearing Correction}

Bandwidth smearing, which is the radio analog of chromatic aberration, arises due to averaging of visibilities over a bandwidth. The effect of the bandwidth smearing on the source is to reduce the peak flux density and to `smear' it around the source; the integrated flux density is unaltered.  Since we make use of the peak flux density of detected sources in the classification of the sources (described in Section \ref{section_source_classification}), in particular to differentiate between unresolved and resolved sources, the bandwidth correction needs to be applied prior to source classification. Below we describe the estimation of the magnitude of this effect.\newline

For the ATLBS survey, the total bandwidth is divided into $2$ bands each with a useful width of $104$ MHz. Each of these bands is further divided into $13$ independent channels of equal width and the data is acquired as multi-channel continuum visibilities, which thus significantly reduces the bandwidth smearing effect. To estimate the bandwidth smearing for the ATLBS sources, we have carried out simulations.  Visibilities corresponding to unresolved sources with a flux density of $1$ mJy were added to data in a single pointing at different distances from the image centre. Any given source in the mosaic is at a maximum distance of $16\farcm4$ from the nearest pointing centre. We utilized the fact that bandwidth smearing depends on the product of the bandwidth and the source distance from the centre and derived estimates of bandwidth smearing within the 8~MHz bands by scaling the offset distances at which sources are added by a factor of $13$ and computing visibilities averaged over a band of $ 13 \times 8$~MHz. The advantage of performing the above procedure is that it does not make any assumptions regarding the shape of the channel bandpass by using data from real observations used for producing the image. A polynomial fit was obtained for the attenuation as a function of distance, which is made use of to correct the measured peak flux densities. The maximum bandwidth smearing correction thus obtained is $8\%$ at a distance of 
$16\farcm4$ from the nearest pointing centre.

The correction factor derived above is correct for unresolved sources.  The peak flux densities for resolved sources are less affected by bandwidth smearing; the correction factors depend on the source structure.  Nevertheless, below we simply apply the derived correction factors to the measured peak flux densities of all sources.  In the case of unresolved sources, the correction is expected to 
aid correct classification; in the case of resolved sources, the overcorrection that results from the above prescription is not expected to alter the classification.

\subsection{On the classification of radio sources in the ATLBS survey}
\label{section_source_classification}

\subsubsection{Initial Source Classification Using the Low Resolution Image}
\label{Initial_classification_subsection}

We find sources and make an initial classification using the low resolution image. The `islands' of pixels above a threshold of 0.38~mJy are identified in the low resolution image and these sky regions are used to estimate the source parameters in the low resolution image.  The sources are classified into (i) unresolved sources, (ii) resolved single-Gaussian sources and (iii) complex sources. A fit with Gaussian models is attempted for all sources. In the case of resolved sources, if the Gaussian model fit is good, then the source is noted as a Gaussian source. If the fit fails and multiple Gaussian are needed to obtain a good fit to the source, then the source is flagged as a multicomponent or complex source. \newline

\subsubsection{Final Source Classification Based on the High Resolution Image}

The final classification of the sources is made using the high resolution image that has beam FWHM of $6\arcsec$, which provides a better representation of the source structure. The high resolution image also helps in distinguishing sources which might be blended together in the low resolution image. For this classification step we use a method similar to that described in the previous section. \newline

Unresolved sources are identified by comparing the peak flux densities in the low and high resolution images. For unresolved sources, we expect that the peak flux at high and low resolution are equal, within errors. If the peak flux in the high resolution image is higher than the peak flux in the low resolution image, we classify the source as unresolved, since this can only happen if noise increases the peak flux in high resolution image and/or lowers it in the low resolution image. If the peak flux in the high resolution image is lower, then we classify the source as a point source if it is within a $2 \sigma$ interval of the low resolution peak. The probability of a noise peak exceeding $2 \sigma$ is $2.2\%$; therefore, approximately $2\%$ of unresolved sources may be misclassified by this scheme. Extended sources (in the high resolution image) might get misclassified as unresolved sources due to noise increasing the high resolution peak flux density or decreasing the peak flux density in the low resolution image so as to satisfy the above condition. However, if the source is extended in the high resolution image, then for sources in which the peak flux density is much greater than the image thermal noise the additive noise required for misclassification is large and has a very small probability associated with it. However, if the source peak flux density is close to the detection threshold and nearer the noise floor, then there is a higher chance that an extended source may get classified as an unresolved source. Therefore, close to the detection threshold, discriminating between unresolved and extended sources might be incorrect. \newline 

The `footprint' of any source is defined to be the area of its `island' in the low resolution image (If the peak flux of the source in the low resolution image is not more than twice the threshold, then the footprint is redefined to be the contour at half the peak flux. This is done so that we search the high resolution image for components in at least a FWHM beam area of the low resolution image). The footprint is used for estimating the flux density and for component identification in the high resolution image. The integrated flux of the source is estimated by iteratively fitting Gaussian models to the source and subtracting these model components till the peak residual in the footprint falls below a threshold. The threshold for any source in a given footprint is determined as outlined in the previous section. 
This may result in underestimation of the integrated flux in the high resolution image when source components in the high resolution image have their peak flux densities below the threshold, but the recipe avoids accumulating spurious flux density from noise peaks to the integrated flux. The number of iterations for each source is noted and the source is assigned a type based on the number of iterations, as discussed below.\newline

If a single iteration is enough to satisfy the above condition then the source, as seen in the high resolution image, is deemed to be a single component. For these sources it is examined if the integrated flux in the low resolution image is the same as the integrated flux in the high resolution image (within errors). Only if both the above conditions are satisfied do we classify the source as a single component Gaussian. Only those sources which have been classified as unresolved or resolved single component Gaussian sources in the low resolution image are examined in the above manner, since any source which is classified as complex in the low resolution image is unlikely to be a single component Gaussian in the high resolution image. The Gaussian sources and the unresolved sources together form the sources whose flux densities have been estimated in the above automated process.  \newline

If multiple iterations are needed for the modeling of the high-resolution intensity distribution within the footprint, or if the sum of the flux densities associated with the components in the high resolution image is less than the integrated flux density measured in the low resolution image, then the source is classified as a complex source.  Multi-component Gaussian modeling of complex sources using the high resolution image may underestimate the integrated flux density of complex sources because such a method might miss flux density associated with extended emission, which may be present in such sources. To avoid such problems, the estimation of integrated flux density for the complex sources is best done not by modeling, but simply summing the pixel intensities enclosed within the `footprint'. Additionally if it is ascertained that the complex source is a single source (even if it has multiple components) and not multiple confused sources, then the integrated flux measured as in the low resolution image may be adopted as the integrated flux density of the source.  \newline

Sources for which the noise gives a fitted diameter less than the beam are considered point sources and the peak flux density is then the optimum estimate for both the flux densities. \newline

For sources classified as unresolved or single-component Gaussians on the basis of the high resolution image, we adopt the low resolution integrated flux density as the estimator of the true integrated flux density, rather than using its counterpart from the high resolution image. It may be noted that the integrated and peak flux densities of the single component Gaussians and unresolved sources are the same, within errors, in the high resolution and low resolution images. \newline

For those sources classified as complex in the high resolution images, we examine whether the source is composed of only a single component in the high resolution image. If the source consists of a single component and the integrated flux contained in the low resolution image is deemed to be the entire integrated flux for the source then we again take the low resolution (fitted) integrated flux as the estimate of the integrated flux density for the source. Those sources that are classified as complex in both the high and low resolution images as well as additional sources declared complex in the high resolution image and displaying multiple components are examined individually by eye rather than by the automated algorithm. \newline

\subsection{The identification of Sources with Complex Structures}

In rare cases, we find that different components of a single source (e.g. lobes of a radio galaxy) lie on separate islands and are hence misidentified as separate sources in the low resolution images. More frequently, we find that unrelated sources are located within a single island in the low resolution image. Each such source was examined by eye ($342$ such sources in all) and it was ascertained if the components composing them are parts of a single source or unrelated sources which
are only close in projection: we adopt the following steps to arrive at a classification. It may be noted here that it is essential to determine the classifications of such sources to better estimate the true source counts and all information available (radio as well as optical) for a given source ought to be used. \newline

1) Optical images (${\rm r}^{\prime}$ band) of the ATLBS survey regions were examined for the presence of hosts for the radio sources. The optical observations in ${\rm r}^{\prime}$ are $90\%$ complete to magnitude of $22.5$ \citep{TSS12}. The host was identified using overlays of the high resolution radio images to identify galaxies in the optical image at the location of the radio core. Where radio cores were undetected, the location of the centroid of the two radio components, which are suspected to be parts of a double source, was examined for the presence of an optical galaxy.  If candidate optical hosts were indeed present at these locations, the pair of radio sources was deemed to be a double radio source and the galaxy was taken to be the host. The optical matching was attempted for $299$ sources, out of which $187$ were found to have optical identifications. Redshift information is available for $24$ of these sources. None of these identifications appeared to be a QSO. \newline

2) The radio sources for which no optical host was found were examined in radio images of intermediate resolution (beam FWHM = 10'') for any connected emission between the radio components. Separately, the integrated flux for these sources was compared to the integrated flux measured in the low resolution image (The estimation of flux density in these cases was done not through model fitting but by summing the pixel intensities within the sky area corresponding to the source). If the integrated flux of the components together is less than the integrated flux as estimated from the low resolution image, then this was interpreted as due to missing extended emission in the high resolution image. Assuming that the missing extended emission would connect the components, it was assumed that the component pair are parts of the same source.\newline

Components deemed to be parts of a single source by the above methods were cataloged as a single source and classified as complex. In those cases where the sources were deemed to be unrelated, the flux for each source was measured separately. Using the high resolution image, the source classification was made as unresolved, single Gaussian or complex and the flux density estimated accordingly for each type (as described in the previous subsection).  

In total, 128 sources in the low resolution image were found to be blends of separate sources. This is ~$10\%$ of the sources found initially in the low resolution image.

Table~\ref{class_table} presents the classification criteria in condensed form. A sample of the ATLBS source catalog is presented in Table~\ref{catalog_table}. The complete catalog is available online.

\begin{table}
\begin{tabular}{|c|c|c|c|}
\hline
 Source  & Classification & Number of & Percentage \\
 Type & Criteria & Sources &  \\
\hline
Unresolved Sources (P) & Peaks flux densities match & 561  & 41.0 \\   
                        & between low and high resolution images. &  & \\
\hline
Single Component  & Single iteration of & 267 & 20.0\\
Gaussian Sources (G) & Gaussian model fitting recovers &  &\\
 & all the integrated flux. &  &\\
\hline
Complex Sources (C) & Multiple iterations & 381 & 28.0 \\
 & of  Gaussian model & & \\
& fitting needed to  &  & \\
& account for all the integrated flux. & & \\
\hline
 
\end{tabular}
\caption{The source types and the classification criteria along with the number of sources of each type and the corresponding percentages.}
\label{class_table} 
\end{table}

\begin{deluxetable}{rrrccc}
\tablewidth{0pt}
\tablecolumns{5}
\tablecaption{The ATLBS survey 1.4 GHz Source Catalog\label{catalog_table}}
\tablehead{
\colhead {Source Name} & 
\colhead {RA (J2000)}  &
\colhead {Dec (J2000)} &
\colhead {$S_{int} $(mJy)}&
\colhead {Source Type}&
\colhead {Number of Components}
}
\startdata
J0025.5-6640	&	00:25:30.82	&	-66:40:31.5	&	1.87	&	G	&		\\
J0025.6-6700	&	00:25:38.90	&	-67:00:19.5	&	1.60	&	G	&		\\
J0025.7-6634	&	00:25:46.04	&	-66:34:35.1	&	0.56	&	M	&		\\
J0025.9-6621	&	00:25:58.71	&	-66:21:20.4	&	215.26	&	C	&	1	\\
J0025.9-6629	&	00:25:54.62	&	-66:29:46.0	&	0.71	&	C	&	1	\\
J0025.9-6632	&	00:25:56.75	&	-66:32:59.7	&	1.89	&	C	&	1	\\
J0025.9-6725	&	00:25:59.89	&	-67:25:38.8	&	1.06	&	P	&		\\
J0026.0-6628	&	00:26:02.54	&	-66:28:06.0	&	0.61	&	M	&		\\
J0026.0-6638	&	00:26:05.38	&	-66:38:15.5	&	0.49	&	C	&	3	\\
J0026.2-6643	&	00:26:12.76	&	-66:43:45.0	&	0.80	&	P	&		\\
J0026.2-6646	&	00:26:12.14	&	-66:46:01.7	&	0.72	&	C	&	1	\\
J0026.3-6708	&	00:26:19.80	&	-67:08:43.0	&	2.36	&	P	&		\\
J0026.3-6713	&	00:26:21.13	&	-67:13:45.7	&	0.69	&	P	&		\\
J0026.4-6721	&	00:26:28.95	&	-67:21:49.7	&	1.68	&	C	&	3	\\
J0026.5-6639	&	00:26:34.15	&	-66:39:53.9	&	1.00	&	G	&		\\
J0026.5-6708	&	00:26:34.95	&	-67:08:14.0	&	5.21	&	G	&		\\
J0026.5-6725	&	00:26:31.40	&	-67:25:04.7	&	3.41	&	P	&		\\
J0026.6-6646	&	00:26:37.80	&	-66:46:03.5	&	2.70	&	G	&		\\
J0026.6-6659	&	00:26:38.46	&	-66:59:35.5	&	0.47	&	P	&		\\
J0026.6-6731	&	00:26:37.16	&	-67:31:06.0	&	1.89	&	G	&		\\
J0026.7-6632	&	00:26:46.00	&	-66:32:00.2	&	5.38	&	G	&		\\
J0026.7-6640	&	00:26:45.80	&	-66:40:06.7	&	0.56	&	P	&		\\
J0026.7-6647	&	00:26:46.38	&	-66:47:38.0	&	0.96	&	C	&	1	\\
J0026.8-6631	&	00:26:48.72	&	-66:31:06.8	&	0.54	&	P	&		\\
J0026.8-6643	&	00:26:49.59	&	-66:43:59.8	&	8.79	&	C	&	3	\\
J0026.8-6649	&	00:26:53.14	&	-66:49:36.4	&	0.99	&	P	&		\\
J0026.8-6659	&	00:26:48.56	&	-66:59:51.3	&	0.45	&	C	&	1	\\
J0026.8-6714	&	00:26:50.49	&	-67:14:03.8	&	0.41	&	M	&		\\
J0026.9-6626	&	00:26:57.67	&	-66:26:18.8	&	1.95	&	G	&		\\
J0026.9-6652	&	00:26:59.90	&	-66:52:14.8	&	0.74	&	C	&	1	\\
J0026.9-6706	&	00:26:55.16	&	-67:06:22.8	&	0.58	&	M	&		\\
J0026.9-6732	&	00:26:56.81	&	-67:32:15.8	&	0.52	&	P	&		\\

\enddata
\tablecomments{The first column gives the name of the source, based on the centroid of the radio emission. The J2000.0 Epoch coordinates of the centroid is given in column 2 and 3. Column 4 lists the integrated flux density. The fifth column gives the type of the source based on the high resolution image. `C' denotes a `complex' source, `G' denotes a single component Gaussian. `P' denotes an unresolved source. `M' denotes a source undetected in the high resolution image. The last column gives the number of components for the complex sources. Table 2 is published in its entirety in the electronic edition of the Astrophysical Journal. A portion is shown here as a sample of its form and content.}
\end{deluxetable}

\section{Source Counts}
\label{section_source_counts}
In this section, we compute the ATLBS source counts and discuss the corrections necessary to estimate the true source counts from the distribution of ATLBS sources in flux density.  First, the sources were binned in octave bins based on their integrated flux density. The effective bin centers need to be defined carefully because in a given bin the source counts are not expected to be uniformly distributed (in which case, the mean would suffice).  The corrections that translate observed counts to true counts depend on this distribution within the bins and, hence, the effective bin center.  The bin centers in our case have been calculated using the expression given by Windhorst et al. (1984):

\begin{equation}
 S_{bc} = [(1-\gamma)\ (\frac{S_{l}-S_{u}}{S_{l}^{1-\gamma}-S_{u}^{1-\gamma}})]^{1-\gamma},
\end{equation}
where $\gamma = 2.52 + 0.321 \times log(S_{gc}) + 0.042 \times log^{2}(S_{gc})$ and $S_{gc}$ is the geometric center of the bin. The factor $\gamma$ has been obtained by a fit to the source counts using a collection of data from \cite{WGK84}, therefore taking care of the source distribution within any bin. We have, additionally, estimated the value of $\gamma$ from the source counts which we have derived; the value of the slope thus estimated is consistent with the above value of the slope over most of the flux density range.\newline

The blending correction to the counts is negligible due to the high angular resolution in our images, and it has been discussed in detail in a previous section.  To assure completeness of the source counts, we have applied resolution bias correction, effective area correction and noise bias correction; these are discussed below.

\subsection{Corrections to the source counts}

\subsubsection{Noise Bias Correction}

Noise bias arises because the `true' source counts are modified by the the noise in the image.  This is true for estimations of any distribution affected by additive noise \citep{Edd13}; noise bias corrections have been applied previously to radio source counts \citep{MCJ73,Ric00,Bon08}.  Noise in the image effectively redistributes sources across neighboring bins, the magnitude of the effect depends both on the image noise and the intrinsic source distribution. If a particular bin has nearby bins with nearly the same numbers of sources, then the number of sources migrating into the bin and out of the bin are the same, assuming symmetric noise distribution. Additionally, at high flux densities, the contribution of  image noise to the source flux density is small and is usually a small fraction of the bin size, which most often increases in geometric progression. Therefore, the effects of noise bias are important for the lowest few bins. \newline

We have simulated the effects of noise bias for our images. We have made use of the polynomial fit to the radio source counts given by Hopkins et al.(2003) as a model for the true source distribution (It may be noted here that later in this paper in Section 5 we derive an improved estimate of noise bias and resultant source counts using  Eqn.~\ref{eqn9}, which is derived from the ATLBS counts for self consistency). The polynomial fit is :

\begin{equation}
 log[(dN/dS)/S^{-2.5}] = \sum\limits_{i=0}^6 \ a_{i}[log(S/mJy)]^{i},
\end{equation}
with $a_{1} = 0.859$, $a_{2} = 0.376$, $a_{3} = -0.049$, $a_{4} = -0.121$, $a_{5} = 0.057$, $a_{6}= -0.008$. 

Noise of random amplitude, consistent with a Gaussian distribution with rms equal to that of the image noise, was added to the integrated flux density of the sources derived from the distribution, and the sources were rebinned. This gives an estimate of the noise bias correction required for each bin. The correction is significant only for the lowest two bins. For the first bin, $0.38-0.768$ mJy, the correction to the source counts is $23.8 \%$ which drops to $3.6 \%$ for the second bin and for the third bin the noise bias correction is only $0.69 \%$. The derived corrections have been applied to the source counts for the above bins since for other bins the correction is negligible.  
 
\subsubsection{Effective Area Correction for Sources}

Only sources that have peak flux densities (without gain correction) above the detection threshold and located in the area of the mosaic where the primary beam gain is above 0.5 are included in deriving the source counts. Approximately $17\%$ sources above the detection threshold lie outside this area. Since the primary beam gain varies over the mosaic image this leads to a bias in source detection because sources with higher peak flux densities are potentially detectable over a larger sky area. Sources with peak flux density above twice the detection threshold may be detected anywhere in the area mentioned above (since the maximum attenuation suffered would lower the peak flux densities above or at the detection threshold); whereas the sources with peak flux densities between the detection threshold and twice that value would be detectable only within a restricted area. The correction factor to account for the change in effective sky area with flux density may be estimated as the ratio of the effective area for a given peak flux density and the total detection area, assuming that sources are uniformly distributed on the sky.  

\subsubsection{Effective Area Correction for Blended Sources}

For blended sources (or rather those sources which are blended in the low resolution image and have been found to be multiple distinct sources at high resolution), the peak flux density as well as the integrated flux density have been determined using the high resolution image. An effective area correction derived from the low resolution peak flux density may be incorrect in these cases because in many such cases sources with a large difference in peak flux density are blended together. Blended sources with individual peak flux densities above the detection threshold would have all been detected irrespective of blending. For these sources we use the high resolution integrated flux density to derive the effective area correction.  Blended sources with peak flux densities below the detection threshold but with integrated flux density above the detection threshold would also have been detected since at low resolution the peak flux density (in most cases) would have been equal to the integrated flux density in the high resolution image. There are $37$ sources with peak flux densities below the detection threshold but with the integrated flux density above the detection threshold. For these sources, therefore, we estimate the effective area correction using the high resolution integrated flux density. Blended sources with integrated and peak flux densities below the detection threshold are not included in the derivation of source counts.

\subsubsection{Resolution Bias Correction}
\label{section_resolution_bias}
Resolution bias is because extended sources that have integrated flux density above the threshold may have their peak flux density below the detection threshold; this would not happen in the case of an unresolved source of the same integrated flux density. Owing to this effect a number of extended sources are lost in the source detection process, biasing the source counts to preferentially represent unresolved sources. We mitigate this effect by using low resolution images for initial source detection. Below we estimate the resolution bias correction applicable to the source counts. We have used the expression given by \citet{WMN90} to represent the angular size distribution of sources and thereby derive the resolution bias. The fraction of radio sources above an angular size $\phi$ is given by :

\begin{equation}
 h(> \phi) = exp\ [\ -ln \ 2 \ (\phi/\phi_{med})^{0.62}],
\end{equation}
where the expression for $\phi_{med}$ (median angular size) is:

\begin{equation}
 \phi_{med} = 2.0" \ S_{1.4 GHz}^{0.3}   
\end{equation}
with $S_{1.4 GHz}$ in mJy.\newline

The relationship between angular size and ratio of integrated to peak flux density is

\begin{equation}
 1 + (\phi/b_{s})^{2}= (S_{int}/S_{peak}).
\end{equation}
For a fixed integrated flux density, a source would be on the threshold of being missed if its peak flux density is equal to the detection threshold; such a source has an angular size:

\begin{equation}
 \phi_{max} = b_{s} \times [(S_{int}/(S_{threshold})) - 1]^{1/2}.
\end{equation}
Here $b_{s} = 50\arcsec$ is the angular size of the synthesized beam for the low resolution images, $S_{threshold}= 0.38$ mJy is the detection threshold and $S_{int}$ is the mean flux density for the source bin. \newline

Therefore, the fraction of sources missing due to being resolved is:

\begin{equation}
  h(> \phi) = exp\ [\ -ln \ 2 \ (\phi_{max}/\phi_{med})^{0.62}].
\end{equation}
Thus the resolution bias correction for each bin is calculated as:
\begin{equation}
 c = 1/(1-h(>\phi_{max})).
\end{equation}
Since the beam FWHM in our case is $50\arcsec$, the resolution bias is only a small correction to the source counts. The maximum effect of the resolution bias is seen in the lowest bin, where an estimated $1.25 \%$ sources are lost due to the resolution bias. In higher flux density bins, the effect of resolution bias is negligible. For source counts generated solely with high resolution images, however, the resolution bias corrections can be significant. 

\subsubsection{The ATLBS source counts}

\begin{figure}
\includegraphics[width = 5 in]{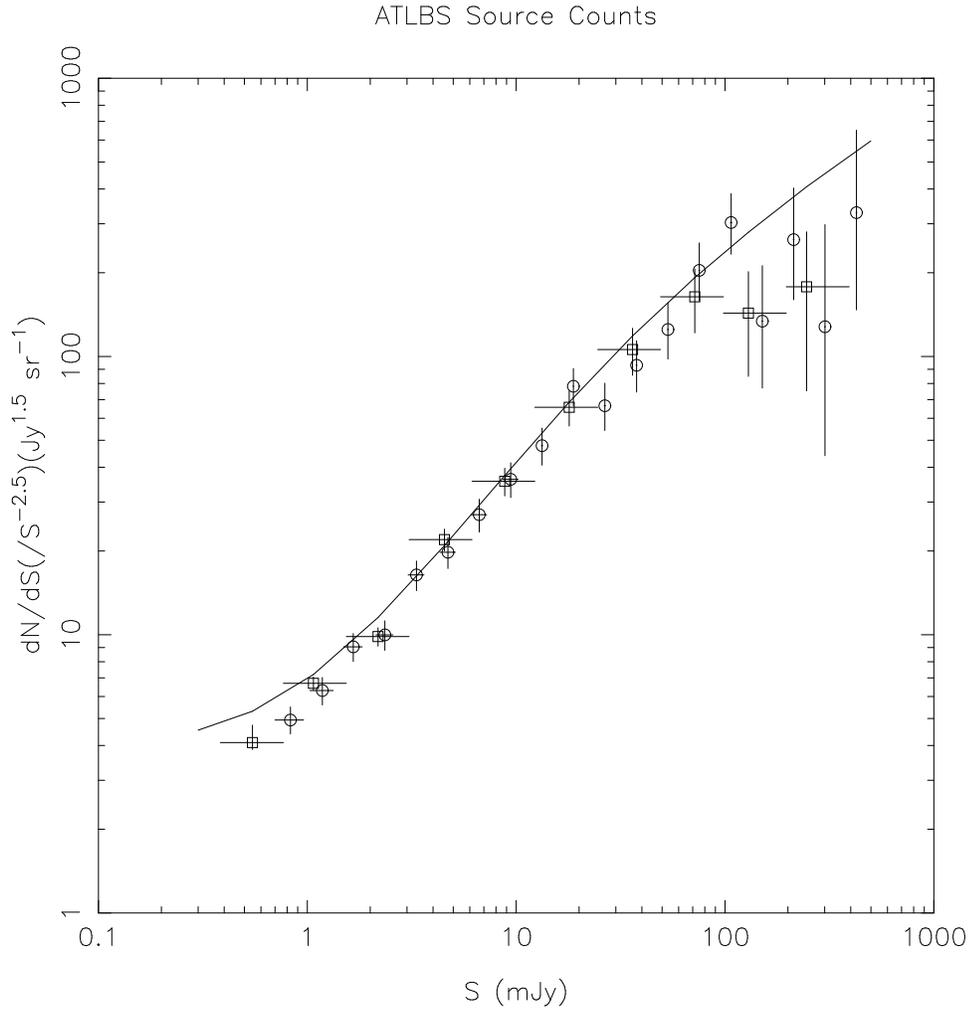}
\caption{The normalized differential source counts for ATLBS are presented (using square symbols with associated error bars). The source counts for ATESP survey \citep{Pra2001} (shown by circles) and the fit to the source counts for ATHDFS survey \citep{HJNP05} (continuous curve) are also depicted for comparison.}
\label{source_counts_figure}
\end{figure}

The total number of sources in the ATLBS survey above a threshold of $0.38$ mJy is $1366$; this is the final number of sources after using the high resolution image to  identify blended sources in the 
low resolution image. The differential source counts, with all of the corrections discussed above, are shown in Fig.~\ref{source_counts_figure}.  The fit to the source counts from the Australia Telescope Hubble Deep Field-South survey (ATHDFS) of \cite{HJNP05} as well as the source counts for the Australia Telescope ESO Slice Project (ATESP) survey of \cite{Pra2001} are overlaid for comparison. Here the fit for ATHDFS is for a compilation of source counts of radio surveys including FIRST (Faint Images of Radio Sky at Twenty-Centimeters; \cite{Beck94}), PDF, ATHDFS and ATESP. There is good agreement between the ATLBS source counts and ATESP source counts. Our source counts are, however, lower than the ATHDFS counts over most of the flux density range by about $~10\%$. The ATLBS source counts derived herein based on high resolution followup of the ATLBS survey is consistent with earlier derivations of ATLBS source counts \citep{SESS10}, which were based on the low resolution images. \newline   
    
\section{A discussion on the slope and magnitude of the counts at sub-mJy flux densities}
\label{section_comparison}

\begin{figure}[H]
\includegraphics[width = 6 in]{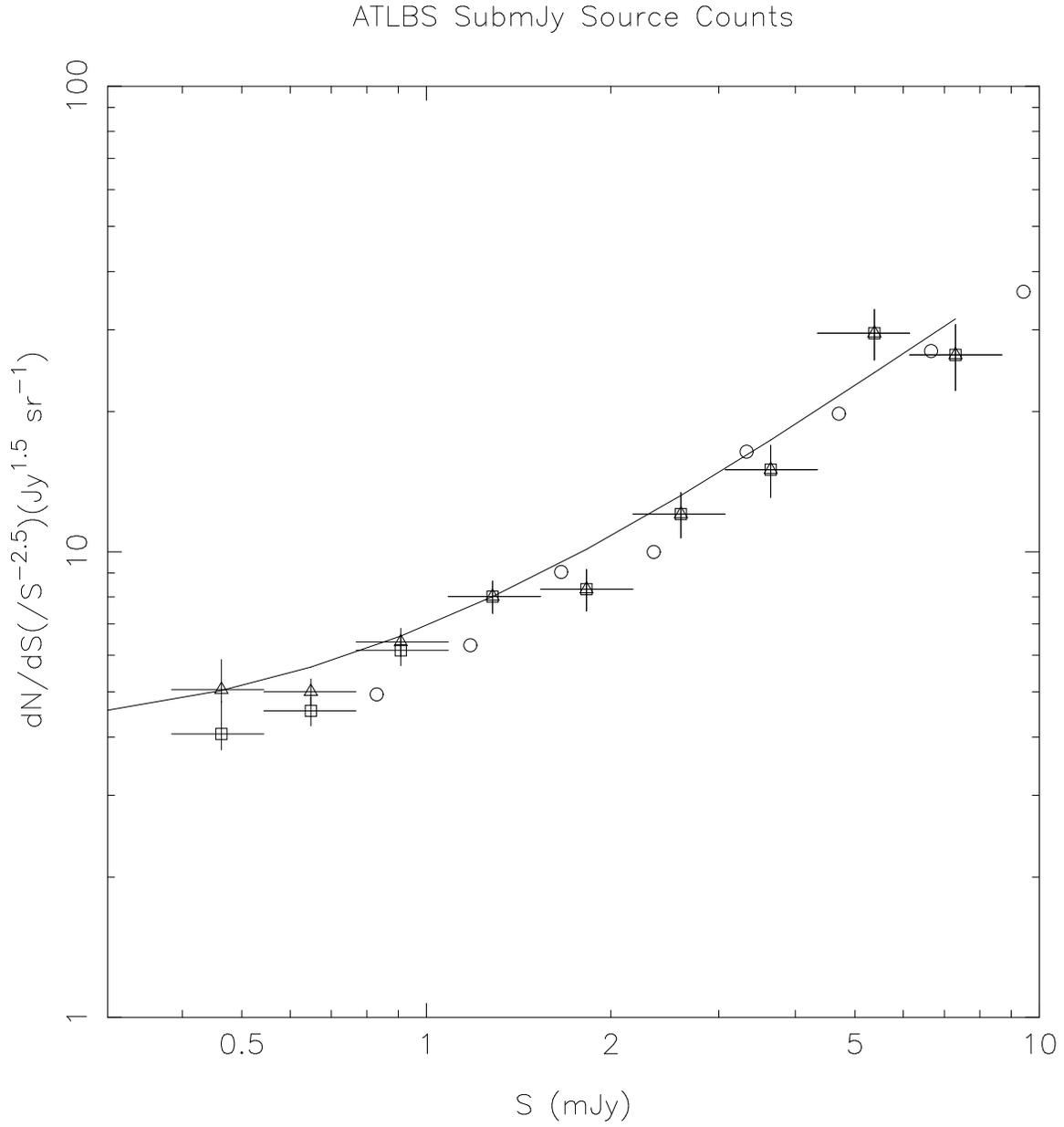}
\caption{The ATLBS differential source counts with reduced bin widths. The squares represent the source counts corrected for noise bias and the triangles represent the source counts without noise bias correction. The other symbols are same as that in Fig.~\ref{source_counts_figure}.}
\label{submil_source_counts}
\end{figure}

The flattening in the normalized differential source counts widely reported in the literature is not obvious in the ATLBS source counts.  To take a closer look at the behavior of the ATLBS counts below about 1~mJy, we have estimated the differential source counts using binning with smaller bin sizes; these are displayed in Fig.~\ref{submil_source_counts}. The ATLBS counts are consistent with the  ATESP counts within the errors and continue the same trend to lower flux densities.   However, the ATLBS counts appear not to follow the upturn of the ATHDFS counts and fall significantly short of the ATHDFS counts below about 1~mJy. \newline

The ATHDFS survey catalog is reported to have been constructed as a source catalog as opposed to a component catalog \citep{HJNP05}. However, the specific forms of the corrections applied to the source counts from the ATLBS and ATHDFS differ, which may explain the difference in the source counts in the two cases. Specifically, we note that the corrections for noise bias are not applied to the source counts for ATHDFS as well as PDF survey \citep{Hop03}. Since the sub-mJy counts are close to the detection threshold, the effect of noise bias is significant and needs to be applied to correctly estimate the source counts. In Fig.~\ref{submil_source_counts} we have displayed the source counts for ATLBS, generated without applying the noise bias correction. It may be seen from the figure that in the absence of noise bias corrections, the differential source counts do exhibit a more pronounced flattening in the sub-mJy regime.  Clearly, omitting noise bias corrections tend to generate flatter estimates for the source counts at levels close to the detection limit, and a factor responsible for the deficit in the ATLBS counts compared to those of the ATHDFS and PDF counts may be their having omitted the noise bias correction. \newline

\begin{figure}
 \includegraphics[width = 6 in]{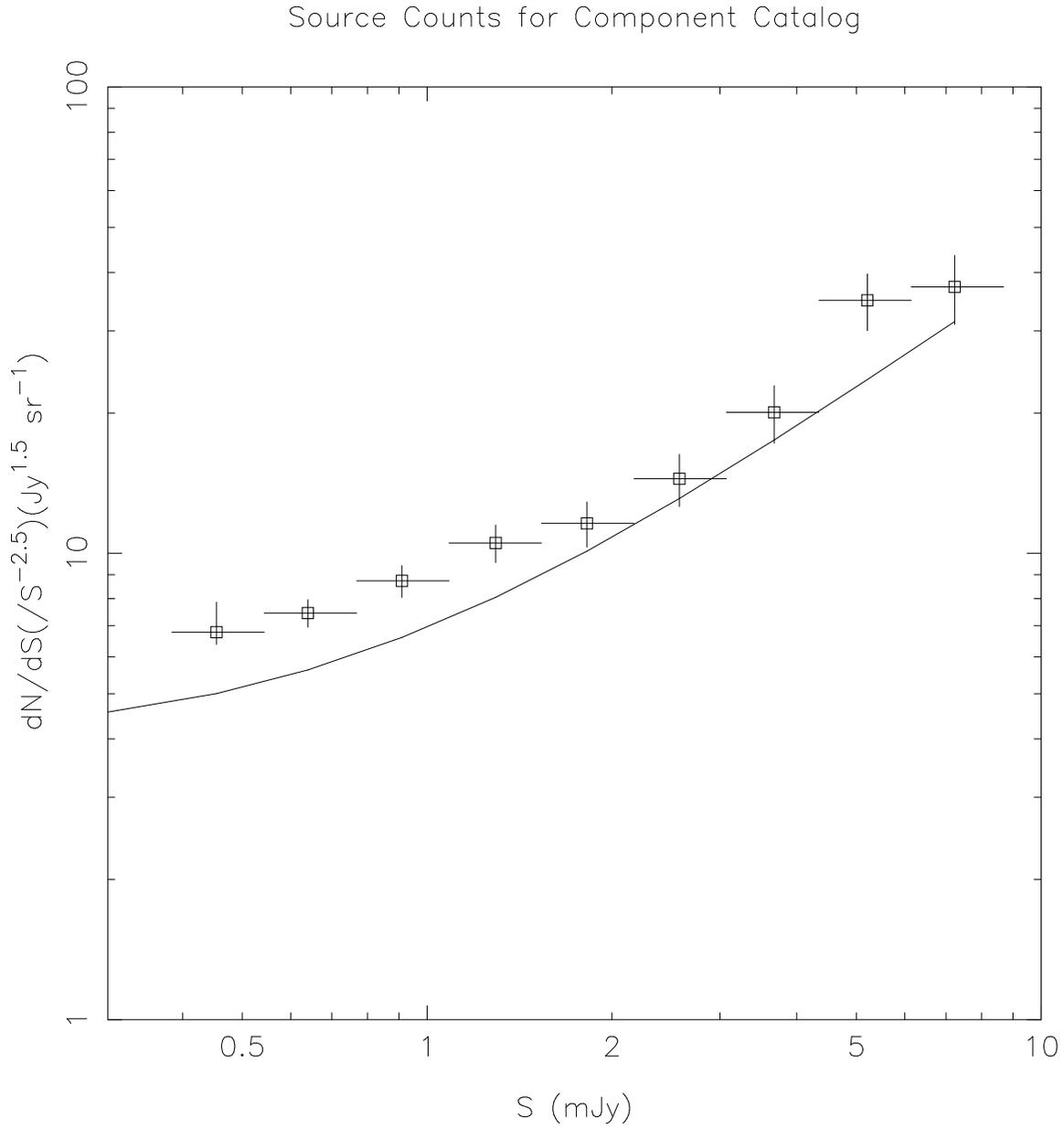}
\caption{The normalized differential source counts for ATLBS, estimated using the high resolution images and a `component' catalog. The continuous curve is the fit from ATHDFS. The sub-mJy counts show a pronounced flattening as well as counts higher than that from ATHDFS.}
\label{submil_source_counts_comp_cat}
\end{figure}

\begin{figure} [H]
 \includegraphics[width = 6 in]{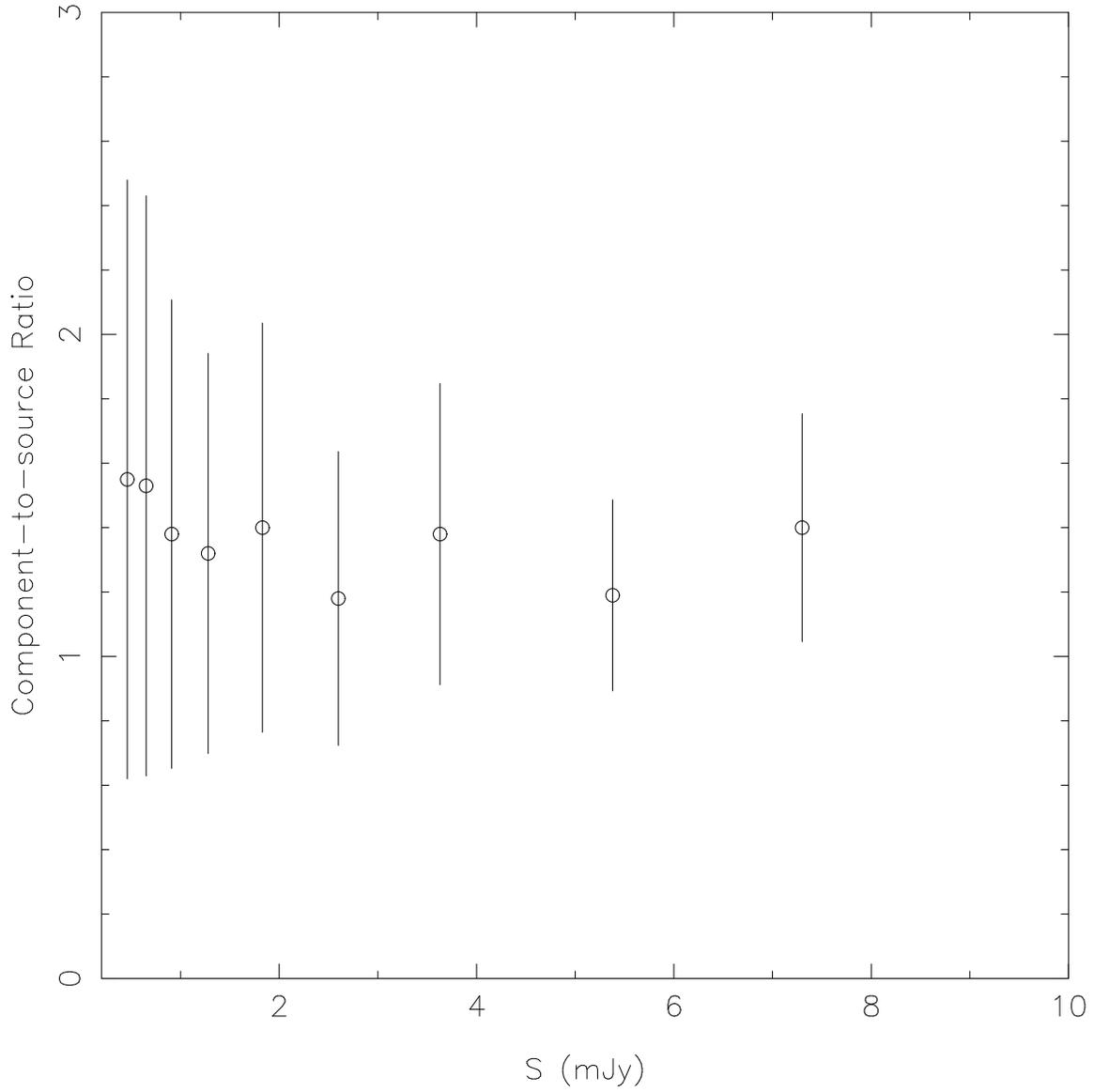}
 \caption{The component-to-source ratio as a function of flux density. Within errors the ratio is constant across the flux density bins.} 
\label{source_to_comp_ratio}
\end{figure}

Source counts derived from `component' catalogs are expected to exhibit extra `sources' at low flux densities, as extended sources are decomposed into components with relatively lower flux densities.  Additionally, in the case of surveys done with high resolution, it is possible for extended sources to break up into two or more compact components if the connecting diffuse emission is missed in the imaging. As a demonstration, we have constructed source counts from a component catalog generated from the ATLBS high resolution images (with beam FWHM of $6\arcsec$). The catalog was generated using the MIRIAD task IMSAD, with a detection threshold of $4 \sigma$, where $\sigma = 72 \ \mu$Jy~beam$^{-1}$ is the rms noise in the image, without any attempt at constructing a `source' catalog as we have done in this work. The source detection was restricted to regions with primary beam gain of $0.9$, so that corrections for the attenuation due to the primary beam, and the effective area correction, may be neglected. The noise bias correction as well as the resolution bias correction (which is important in the case of images with high resolution) were separately derived for these counts, in the same manner as described in earlier sections. In Fig.~\ref{submil_source_counts_comp_cat} we show the source counts generated from this component catalog. As may be expected, this results in a substantially flatter distribution for the source counts; additionally,  the source counts are now greater than before as well as greater than the ATHDFS counts. In Fig.~\ref{source_to_comp_ratio} we display the ratio of component counts to source counts versus flux density. This ratio has an average value of 1.4 implying that on average the component counts are a factor 1.4 higher than the true source counts.\newline

The ATLBS counts are about $10\%$ lower than the ATHDFS counts over the 0.4-10~mJy range.  As seen in Fig.~\ref{submil_source_counts_comp_cat}, a component catalog may overestimate the counts in this flux density range by as much as $30-50\%$; \citet{Hop03} estimate that a component catalog may overestimate the counts by about $10\%$.  The quantum of error would depend on the structures of radio sources at these flux densities and the quality of the imaging---the ability of the imaging to reproduce any connecting emission between components---and hence the correction factor necessary for deriving a source count from component counts may vary depending on the visibility coverage and imaging algorithms.  Nevertheless, we conclude that the relatively lower source counts inferred in this work compared with, for example, PDS and ATHDFS counts, are in part owing to the ATLBS counts having been carefully prepared to represent a source catalog rather than a component catalog. Note that catalogues such as NVSS are component and not source catalogues so any source count analysis based on these catalogues will have this problem. It is possibly for this reason that no source count
analysis is included in the NVSS publication \citep{Con98}.\newline

The noise bias correction depends on the assumption of what is the true source counts: this prior assumption is used to determine the true number of sources in each bin so that the effect of noise bias may be estimated. As discussed earlier, we have adopted the
PDF source counts for the determination of the noise bias. However, the PDF source counts show a more pronounced flattening compared to the ATLBS; therefore, we have carried out a second iteration of noise bias correction using the derived ATLBS  source counts. 

We fitted a second order polynomial in flux density to the ATLBS source counts derived above, which appears sufficient to express the features seen in our source counts. The polynomial fit is given by :

\begin{equation}
 log[(dN/dS)/S^{-2.5}] = 0.781 + 0.851\times log(S/mJy)-0.066\times log^{2}(S/mJy). 
\label{eqn9}
 \end{equation}
 
 Since the ATLBS source counts are relatively steeper compared to the PDF counts, the noise bias correction estimated from the ATLBS counts is smaller. For the lowest two bins 0.38-0.54 and 0.54-0.77~mJy, the noise bias correction is now derived to be $11.18\%$ and $5.22\%$ respectively, reduced from the previous values of $24.5\%$ and $10\%$. This improved estimate of the source counts is shown in Fig.~\ref{submil_source_counts_new} and Table~(\ref{submil_counts_table_new}). The ATLBS source counts are still systematically lower than the ATHDFS source counts; however, the difference between the two source counts is reduced.   

\placetable{submil_counts_table_new}
\begin{deluxetable}{rrccc}
\tablecaption{ATLBS 1.4 GHz source counts with a self-calibrated noise bias\label{submil_counts_table_new}}
\tablewidth{0pt}
\tablecolumns{5}
\tablehead{
\colhead{$\Delta S$ mJy} &
\colhead{$\langle S \rangle$ mJy } &
\colhead{$dN/dS (/S_{-2.5})(Jy^{1.5}Sr^{-1})$} &
\colhead{$(dN/dS)_{pdfnb}$ } &
\colhead{$dN/dS_{wonb}$ } 
}
\startdata
$	0.38	  -	0.54	$	&	$	0.46	$	&	$	4.55	(+0.76,-0.31)	$  &$ 4.06 (+0.71,-0.30)$   & $5.05 (+0.81,-0.31)$	\\
$	0.54	  -	0.77	$	&	$	0.65	$	&	$	4.76	\pm	0.33	$ &$4.55 \pm 0.33$  &$5.00 \pm 0.33$	\\
$	0.77	  -	1.09	$	&	$	0.91	$	&	$	6.20	\pm	0.45	$ &$6.14 \pm 0.45$   &$6.40 \pm 0.44$	\\
$	1.09	  -	1.54	$	&	$	1.28	$	&	$	7.90	\pm	0.64	$ &$8.01 \pm 0.64$   &	$8.02 \pm 0.64$\\
$	1.54	  -	2.17	$	&	$	1.83	$	&	$	8.25	\pm	0.85	$ &$8.31 \pm 0.85$   &	$8.32 \pm 0.85$\\
$	2.17	  -	3.07	$	&	$	2.60	$	&	$	12.05	\pm	1.35	$ &$12.05 \pm 1.34$   &	$12.05 \pm 1.35$\\
$	3.07	  -	4.34	$	&	$	3.63	$	&	$	15.00	\pm	1.92	$ &$15.01 \pm 1.92$    &$15.01 \pm 1.92 $	\\
$	4.34	  -	6.14	$	&	$	5.38	$	&	$	29.48	\pm	3.69	$ &$29.48 \pm 3.69$     & $29.49 \pm 3.69 $	\\
$	6.14	  -	8.69	$	&	$	7.30	$	&	$	26.49	\pm	4.30	$ &$26.49 \pm 4.30$    & $26.49 \pm 4.30$	\\
\hline
											
\enddata
\tablecomments{The table above gives the source counts for the ATLBS survey. The third column gives the source counts with self-calibrated noise bias. The fourth column gives the source counts with noise bias derived from PDF source counts, while the last column has the source counts without noise bias correction.}
\end{deluxetable}

\section{Summary and conclusions}
\label{section_conclusions}

High-resolution radio images of the ATLBS survey regions are presented, with beam FWHM of $6\arcsec$.  The wide field mosaic images covering $8.42$ square degrees sky area with rms noise $72\ \mu$Jy~beam$^{-1}$ are of exceptional quality in that there are no imaging errors or artifacts above the thermal noise over the entire field of view.  The images have excellent surface brightness sensitivity - the visibility coverage is complete out to 750~m and hence provides good representation of extended emission components associated with radio sources.  The images are, therefore, an excellent resource for examining with automated algorithms for source finding, parameter fitting, and morphological classification, and as a resource for testing such algorithms that would be used on upcoming all-sky continuum surveys with the LOFAR and ASKAP.  We make the high-resolution ATLBS images available at the website www.rri.res.in/~ATLBS.

We have generated a source list from the ATLBS images.  This is a carefully made source list as opposed to a component list.  The images were initially examined using automated algorithms, which used representations with different resolutions, to identify sources and distinguish unresolved and single component sources and complex sources.  All complex sources were carefully examined by eye to recognize blends and classify appropriately.  Optical surveys of the ATLBS fields were also examined for candidate host galaxies  to aid in the classification of complex sources.   The integrated flux densities of the sources were derived in a variety of methods - the method appropriate for the source structural classification was adopted in each case. We emphasize the use of multi-resolution images, which may complement each other, as well as the need to use data from other wavebands such as optical, infrared etc. The source list is also presented online along with the integrated flux densities and classification.

\begin{figure}[H]
 \includegraphics[width = 6 in]{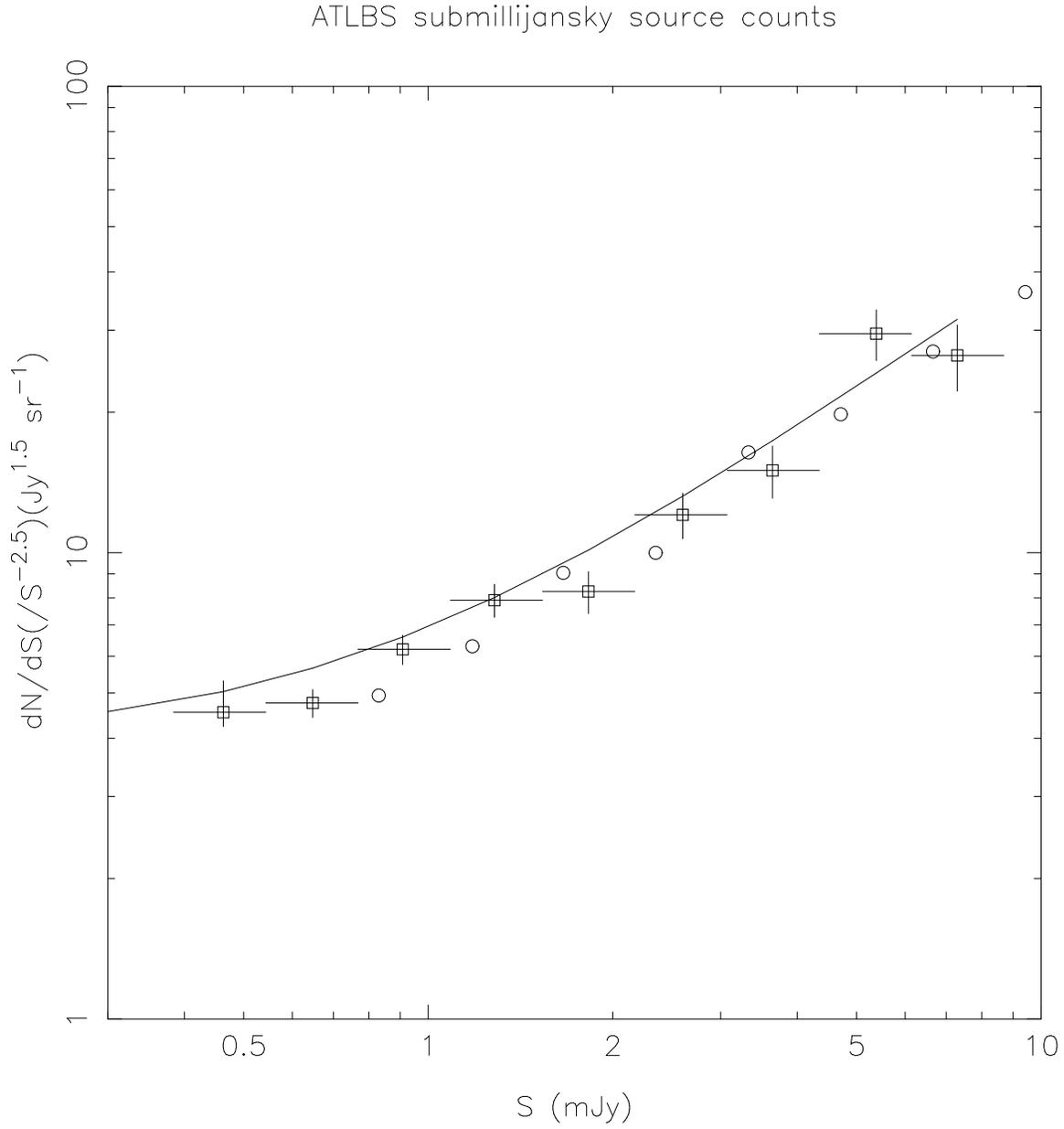}
\caption{The differential source counts for ATLBS derived using noise bias correction derived from the ATLBS counts themselves. The other symbols are same as that in Fig.~\ref{source_counts_figure}.}
\label{submil_source_counts_new}
\end{figure}

The source list was used to estimate the radio source counts down to 0.4~mJy.  The counts have been corrected for noise bias, resolution bias, and effective area.  It may be noted that considerable care has been taken to ensure that the counts correspond to sources and not components.  The counts presented in Fig.~\ref{submil_source_counts_new} and Table~(\ref{submil_counts_table_new}) above have been self-calibrated for the noise bias in that the counts derived in a first iteration have been used to derive the noise bias correction.  

Comparing the counts with previous work - the ATHDFS and PDS counts - shows that the ATLBS counts are systematically lower.  This is attributed to our counts representing sources as opposed to components, as well as corrections for noise bias.  We have demonstrated the substantial difference in counts that results from using component catalogs as opposed to source catalogs: at 1~mJy flux density component counts may be as much as 50\% above true source counts.  This implies that automated image analysis for counts may be dependent on the ability of the imaging to reproduce connecting emission with low surface brightness and on the ability of the algorithm to recognize sources, which may require that source finding algorithms effectively work with multi-resolution and multi-wavelength data. The work presented herein underscores the importance of noise bias correction, in particular for deriving counts close to the limit of the survey sensitivity and for correctly estimating the faint end slope and upturn in the source counts at sub-mJy flux density. Finally, the lack of an upturn in the source counts at faint flux densities implies that down to the faintest flux densities we have probed (approximately 0.4 mJy) there is no evidence for any new population. The upturn reported in the literature may be due to a combination of small survey areas and identification of radio sources as opposed to components, as well as effects of noise bias close to the detection limit.  

\acknowledgements
The ATCA is part of the Australia Telescope, which is funded by the Commonwealth of Australia for operation as a National Facility managed by CSIRO.  We thank Anant Tanna for his assistance with the initial processing of the visibility data.

\newpage

\newpage

\begin{thebibliography}{}
\bibitem[Becker et al.(1994)]{Beck94}Becker, R. H., White, R. L., \& Helfand, D. J. 1994, ASPC, 61, 165 
\bibitem[Benn et al.(1993)]{Ben93}Benn, C. R., Rowan-Robinson, M., McMahon, R. G., Broadhurst, T. J., \& Lawrence, A. 1993, MNRAS, 263, 98
\bibitem[Bondi et al.(2008)]{Bon08} Bondi, M., Ciliegi, P., Schinnerer, E., et al. 2008, Astrophys. J., 681, 1129
\bibitem[Clark(1980)]{clark} Clark, B.G. \ 1980 \ A\&A, Vol. 89, No. 3, 377-378
\bibitem[Condon(1989)]{Con89} Condon, J. J., 1989, ApJ, 338, 13
\bibitem[Condon et al.(1998)]{Con98}Condon, J. J., Cotton, W. D., Greisen, E. W., Yin, Q. F., Perley, R. A., Taylor, G. B., and Broderick, J. J. 1998, AJ, 115, 1693
\bibitem[Eddington(1913)]{Edd13} Eddington, A., 1913, MNRAS, 73, 359
\bibitem[Gruppioni et al.(1999)]{GMZ99}Gruppioni, C., Mignoli, M., \& Zamorani, G. 1999, MNRAS, 304, 199
\bibitem[Hopkins et al.(2003)]{Hop03} Hopkins, A. M., Afonso, J., Chan, B., et al. 2003, AJ, 125, 465
\bibitem[Huynh et al.(2005)]{HJNP05} Huynh, M. T., Jackson, C. A., Norris, R. P., \& Prandoni, I. 2005, AJ, 130, 1373
\bibitem[Huynh et al.(2008)]{HJNF08} Huynh, M. T., Jackson, C. A., Norris, R. P., \& Fernandez-Soto, A. 2008, AJ, 135, 2470
\bibitem[Jauncey(1968)]{Jaun68} Jauncey, D. L., 1968, ApJ, 152, 647
\bibitem[Mitchell \& Condon(1985)]{MC85} Mitchell, K. J., Condon, J. J. 1985, AJ, 90, 1957
\bibitem[Murdoch(1973)]{MCJ73} Murdoch, H.S.,Crawford, D.F., Jauncey, R.L., 1917, ApJ.,118,1
\bibitem[Prandoni et al.(2001)]{Pra2001}Prandoni, I., Gregorini, L., Parma, P., de Ruiter, H. R., Vettolani, G., Wieringa, M. H., Ekers, R. D., 2001, A\&A, 365, 392-399
\bibitem[Richards(2000)]{Ric00} Richards, E.A., 2000, AJ, 533, 611
\bibitem[Saripalli et al.(2011)] {SST11} Saripalli, L., Subrahmanyan, R., Thorat, K., Ekers, R.D., Hunstead, R.W., Johnston,
H.M., Sadler, E.M., 2011, ApJ., In Press
\bibitem[Sault et al.(1995)]{STW95} Sault, R.J., Tueben, P.~J. and Wright, M.~C.~H. 1995, Astronomical Data Analysis Software and Systems IV, ASP Conference Series, Vol. 77, ed. R.A. Shaw, H.E. Payne, and J.J.E. Hayes, 433 
\bibitem[Sault \& Wieringa(1994)]{sault} Sault, R.J. \& Wieringa, M.H.\ 1994 \ A\&AS, 108, 585 
\bibitem[Subrahmanyan et al.(2010)]{SESS10} Subrahmanyan, R., Ekers, R. D., Saripalli, L., Sadler, E. M., 2010, MNRAS, 402, 2792
\bibitem[Thorat et al.(2012)]{TSS12} Thorat, K., Saripalli, L., Subrahmanyan, R., 2012, In Preparation.
\bibitem[Windhorst et al.(1984)]{WGK84} Windhorst, R. A., van Heerde, G. M., \& Katgert, P., 1984, A\&AS, 58, 1
\bibitem[Windhorst et al.(1985)]{WMO85} Windhorst, R.A., Miley, G.K., Owen, F.N., Kron, R.G. and Koo, D.C.,1985, AJ, 289, 494 
\bibitem[Windhorst et al.(1990)]{WMN90}Windhorst, R.A., Mathis, D., Neuschaefer, L.,1990, PASP, Vol. 10, Evolution of the Universe of Galaxies, ed. Kron, R.J., 389-403
\end{thebibliography}
\end{document}